\documentclass[a4paper,11pt]{article}
\usepackage{latexsym}
\usepackage{cite}
\usepackage{epsfig}
\usepackage{amssymb}
\usepackage{latexsym}
\usepackage{epsfig, subfigure, subfloat, graphicx, float}
\usepackage{mathrsfs}
\usepackage{amsmath}
\usepackage{color}
\usepackage[usenames]{xcolor}
\graphicspath{{Image/}}
\author{H. Mohseni Sadjadi\footnote{mohsenisad@ut.ac.ir}  and M. Khodaei\footnote{mahdi.khodaei@ut.ac.ir}
\\ {\small Department of Physics, University of Tehran,}
\\ {\small P. O. B. 14395-547, Tehran 14399-55961, Iran}}
\title{No scalar hair in nonminimal derivative coupling model for Neumann and Dirichlet reflecting star}
\begin{document}
\maketitle
\begin{abstract}
We investigate the existence of scalar hair in the background of a reflecting star. The kinetic part of the scalar field is nonminimally coupled to the Einstein tensor. For both the Dirichlet and the Neumann boundary conditions, we study the no-hair theorem. For the massless case, we show that there is no scalar hair, and in the massive case conditions leading to the no-hair theorem are specified.
\end{abstract}

\section{Introduction}
"No-hair theorem" \cite{nohair1,nohair2,radu} states that a black hole is completely specified by its mass, electric charge and angular momentum and characteristics of an object other than these three ones disappear when it crosses the horizon. Outside the horizon, the fields originated from the black hole are only Maxwell and gravitational fields corresponding to the Gauss law.

Scalar fields, do not obey the Gauss law, and a black hole has not a canonical scalar hair minimally coupled to gravity. The answer to the question "Why should scalar fields be different from electromagnetic fields" was discussed in detail in \cite{radu}, by comparing electro-vacuum with scalar-vacuum in the general relativity.

To see if the no-hair theorem for canonical scalar fields is the exclusive property of black-holes, this theorem was investigated also for a compact star with a Dirichlet boundary condition in \cite{hod1}. Instead of the absorbing event horizon, the boundary acts as a reflecting (repulsing) surface. This star is dubbed "reflecting star". In \cite{hod1} it is shown that as the black hole case, there is no regular nontrivial canonical massive scalar field outside the boundary. No scalar hair theorem with the Dirichlet ($\psi(r_{s})=0$), and Neumann ($\psi^{\prime}(r_{s})=0$) boundary conditions on the boundary surface (located at $r_s$) has been discussed in the literature (see \cite{hod1,hod2,hod3,hod4}, \cite{sark}, and \cite{peng1,peng2,peng3,peng4,peng5,paper1,paper2}).

 Modifying the theory of gravity, or considering non-minimal couplings of a field to the gravity, may alter the situation such that the black hole gains hair \cite{radu}. Non-minimal coupling of the scalar field to the Ricci scalar leads to a non-primitive scalar hair around the black hole \cite{beck1,beck2}, although this scalar field solution is not regular at the horizon of the black hole.  Another example of non-minimal couplings is the non-minimal derivative coupling model, where the kinetic part of the scalar field is coupled to the Einstein tensor. Initially, this model was introduced to describe inflation \cite{inf1,inf2,inf3,inf4,inf5,inf6,inf7}, and the late time acceleration \cite{late0,late00,late1,late2,late3,late4,late5,late6,late7,late8,late9,late10}. The wormhole solution of this model was studied in \cite{Sushkov1,Sushkov2}. The black hole solution was investigated in \cite{Masato,b1,b2,r1,r2}, and also the stability and odd-parity perturbation of this model is discussed in \cite{Adolfo}. There is a brief discussion about the existence of the scalar hair in non-minimal derivative coupling in \cite{Kolyvaris}.

Motivated by the above discussions, we aim to investigate the existence of a scalar hair for a reflecting star in the non-minimal derivative coupling model.
In the second section, we introduce the model and derive the equations of motion. We then try to obtain conditions required for the validity of no scalar hair theorem with both the Dirichlet and Neumann boundary conditions.

We use units  $G=c=\hbar=1$ and the metric signature is $(-+++)$.
\section{The model}
The model with non-minimal derivative coupling of a scalar field to the Einstein tensor is described by the action
\begin{equation}\label{1}
S=\int d^{4}x\sqrt{-g}\left(\frac{R}{8\pi}-[ g_{\mu\nu}+kG_{\mu\nu}]\psi ^{,\mu}\psi ^{,\nu}-2V(\psi)\right).
\end{equation}
Where $k$ is a constant with dimension of length-squared, $\psi$ is the scalar field, $G_{\mu\nu}$ is the Einstein tensor, and $V(\psi)$ is the scalar field potential. One can obtain the Einstein equation by taking the variation of this action with respect to the metric $g_{\mu\nu}$\cite{Sushkov1}
\begin{equation}\label{2}
G_{\mu\nu}=8\pi[ T_{\mu\nu}+k \Theta _{\mu\nu}]-8\pi g_{\mu\nu}V(\psi),
\end{equation}
where $T_{\mu\nu}$ is given by
\begin{equation}\label{3}
T_{\mu\nu}=\nabla _{\mu}\psi\nabla _{\nu}\psi -\frac{1}{2}g_{\mu\nu}(\nabla _{\alpha}\psi\nabla^{\alpha}\psi).
\end{equation}
$\Theta_{\mu\nu}$ is resulted from the derivative coupling of the scalar field to the Einstein tensor and is given by
 \begin{eqnarray}\label{4}
\Theta _{\mu\nu}&=&-\frac{1}{2}\nabla _{\mu}\psi\nabla _{\nu}\psi R+2\nabla _{\alpha}\psi\nabla _{(\nu}\psi R^{\alpha}_{\nu)}\nonumber\\
&&+\nabla ^{\alpha}\psi\nabla ^{\beta}\psi R_{\mu\alpha\nu\beta}+\nabla _{\mu}\nabla ^{\alpha}\psi\nabla _{\nu}\nabla _{\alpha}\psi\nonumber\\
&&-\nabla _{\mu}\nabla _{\nu}\psi\square\psi -\dfrac{1}{2}(\nabla _{\alpha}\psi\nabla^{\alpha}\psi)G_{\mu\nu}\nonumber\\
&&+g_{\mu\nu}[-\frac{1}{2}\nabla ^{\alpha}\nabla ^{\beta}\psi\nabla _{\alpha}\nabla _{\beta}\psi +\frac{1}{2}(\square \psi)^{2}\nonumber\\
&&\qquad\quad -\nabla _{\alpha}\psi\nabla _{\beta}\psi R^{\alpha\beta}].
\end{eqnarray}
 Also variation of the action (\ref{1}) with respect to the scalar field leads to the scalar field equation of motion which is given by
 \begin{equation}\label{5}
[ g^{\mu\nu}+kG^{\mu\nu}]\nabla _{\mu}\nabla _{\nu}\psi =V_{\psi},
\end{equation}
where $V_{\psi}$ is the derivative of potential function with respect to the scalar field $\psi$.

Now consider the space-time outside of the horizon of a black hole. Multiplying both sides of (\ref{5}) by $\psi$ and integrating over space-time, provided that the boundary term vanishes, gives
\begin{equation}
\int \left(g^{\mu \nu }+k G^{\mu \nu}\right)\nabla_\mu \psi \nabla_\nu \psi \sqrt{-g}d^4x+\int \psi V_\psi \sqrt{-g}d^4x=0.
\end{equation}
For $k=0$, we recover the result obtained in the minimal case, implying that for $\psi V_\psi\geq 0$, we have no a nontrivial scalar field, confirming no scalar hair theorem. But for $k\neq 0$, one can evade no-hair theorem as verified in \cite{Masato}. Is this the same for a reflecting star?  to answer this question, let us consider field equations \eqref{2} and \eqref{5} for an asymptotically flat static reflecting star.  We choose the spherically symmetric metric
\begin{equation}\label{6}
ds^{2}=-N(r)A(r)dt^{2}+A^{-1}(r)dr^{2}+r^{2}d\Omega ^{2},
\end{equation}
where $N(r)$ and $A(r)$ are functions of the radial coordinate and  $N(r)\neq 0$, $A(r)\neq 0$  in the region under study. Asymptotical flatness requires
\begin{equation}
\lim_{r\rightarrow\infty} A(r)=1,
\end{equation}
and
\begin{equation}
\lim_{r\rightarrow\infty} N(r)=1.
\end{equation}

With the metric (\ref{6}), the Einstein equations \eqref{2} become
\begin{align}\label{7-1}
&\frac{A^{\prime}}{rA}+\frac{1}{r^2}-\frac{1}{Ar^2}=\nonumber\\
&-4\pi \psi^{\prime 2}-8\pi A^{-1}V(\psi)+8\pi k\psi^{\prime 2}[\frac{3A^{\prime}}{2r}+\frac{A}{2r^{2}}+\frac{1}{2r^{2}}]+16\pi k \frac{A}{r}\psi^{\prime}\psi^{\prime\prime},
\end{align}
\begin{align}\label{7-2}
&\frac{A^{\prime}}{Ar}+\frac{1}{r^{2}}+\frac{N^{\prime}}{rN}-\frac{1}{Ar^{2}}=\nonumber\\
&8\pi k\psi^{\prime 2}[\frac{3A^{\prime}}{2r}+\frac{3A}{2r^{2}}+\frac{3AN^{\prime}}{2Nr}-\frac{1}{2r^{2}}]-8\pi A^{-1}V(\psi)+4\pi\psi^{\prime 2},
\end{align}
\begin{align}\label{7-3}
&2N\frac{A^{\prime\prime}}{A}+2N^{\prime\prime}+3\frac{N^{\prime}A^{\prime}}{A}+2\frac{N^{\prime}}{r}-\frac{N^{\prime 2}}{N}+4\frac{NA^{\prime}}{Ar}=\nonumber\\
&-16\pi N \psi^{\prime 2}-32\pi A^{-1}NV(\psi)+8\pi k\psi^{\prime }\psi^{\prime \prime}[2NA^{\prime }+2AN^{\prime }+4N\frac{A}{r}]\nonumber\\
&+8\pi k \psi^{\prime 2}[A^{\prime \prime}N+AN^{\prime \prime}+N\frac{A^{\prime 2}}{A}-\frac{A}{2}\frac{N^{\prime 2}}{N}+\frac{5}{2}A^{\prime}N^{\prime}+N^{\prime}\frac{A}{r}+4N\frac{A^{\prime}}{r}],
\end{align}
and also the scalar field equation \eqref{5} reduces to
\begin{align}\label{7-4}
&A[kr(A^{\prime}+A\frac{N^{\prime}}{N})+r^{2}+k(A-1)]\psi^{\prime\prime}\nonumber\\
&+krA[A^{\prime\prime}
+A\frac{N^{\prime\prime}}{N}+\frac{5}{2}\frac{N^{\prime}}{N}A^{\prime}-A\frac{N^{\prime 2}}{2N^{2}}]\psi^{\prime}\nonumber\\
&+[(r^{2}+3kA-k )A^{\prime}+(r^{2}+3kA-k)A\frac{N^{\prime}}{2N}]\psi^{\prime}\nonumber\\
&+[krA^{\prime 2}+2rA]\psi^{\prime}=r^{2}V_{\psi}.
\end{align}
A "prime" denotes derivatives with respect to the radial coordinate $r$. For large $r$, i.e. $r\gg r_s$, the scalar field equation\eqref{7-4} asymptotically  becomes
 \begin{equation}\label{9}
 \psi ^{\prime\prime}+\frac{2}{r}\psi ^{\prime} =V_{\psi},
 \end{equation}
where $V_{\psi}=\frac{dV}{d\psi}$.  Equations \eqref{7-1}-\eqref{7-4} are a system of four nonlinear differential equations whose the possible analytical solutions are hard to find. In the following we investigate the possibility to find a non-trivial continuous well defined solution for the scalar field in the region bounded by a reflecting surface and infinity: $r\in [r_s,\infty)$.  We first consider the free massless scalar field  $V=0$, whose wormhole and black holes solutions in nonminimal derivative coupling model were studied in {\cite{Masato} and \cite{Sushkov1} respectively. We show that there are not non trivial solution when the Neumann boundary condition is considered, and then generalize our result to a massive scalar field by a different method.

\subsection{No scalar hair for $V(\psi)=0$}
In this case, (\ref{7-1})and (\ref{7-2}) reduce to:
\begin{equation}\label{s1}
r A'+A-1=-4\pi r^2A\psi'^2+8\pi k\psi'^2\left(\frac{3}{2}A'Ar+\frac{1}{2}A^2+\frac{1}{2}A\right)+16\pi krA^2\psi'\psi'',
\end{equation}
and
\begin{equation}\label{s2}
r A'+A-1+\frac{N'}{N}Ar=8\pi k \psi'^2\left(\frac{3}{2}A'Ar+\frac{3}{2}A^2+\frac{3N'}{2N}A^2-\frac{1}{2}A\right)+4\pi r \psi'^2A^2
\end{equation}
respectively. By using Neumann boundary condition at $r=r_s$ we obtain
\begin{equation}\label{n1}
(rA'+A)\big|_{r_s}=1
\end{equation}
and
\begin{equation}\label{n2}
(rA'+A+\frac{N'}{N}Ar)\big|_{r_s}=1
\end{equation}
resulting $N'(r_s)=0$. We now show that all higher derivative of $\psi$ at $r_s$ are also zero. From (\ref{7-4}) we have $\psi''(r_s)=0$. Now by taking consecutive radial derivative of (\ref{7-4}) at $r_s$ we find that at each derivative order $j$
\begin{equation}
\left(A[kr(A^{\prime}+A\frac{N^{\prime}}{N})+r^{2}+k(A-1)]\psi^{(j)}\right)\big|_{r_s}=0
\end{equation}
But from (\ref{n2}) we have $\left(A[kr(A^{\prime}+A\frac{N^{\prime}}{N})+r^{2}+k(A-1)]\right)\big|_{r_s}=A(r_s)r_s^2$, therefore $\psi^{(j)}(r_0)=0$.
So the only analytic well defined $\psi(r)$ is a trivial constant function, $\psi(r)=0$.

If instead, one considers Dirichlet boundary condition at $r_s$; $\psi(r_s)=0$, and assumes that the scalar field tends to zero at infinity $(\psi (r)\propto
\frac{2}{r})$, then there must be a point $r_0$ where $\psi'(r_0)=0$. By applying the above argument we find all radial derivatives of $\psi$ vanishes at $r_0$, and therefore $\psi(r)$  must be a constant function with the same value at the infinity and on the reflecting surface.

\subsection{ Massive scalar field and the no hair theorem}
In this part, we consider a mass for the scalar field
\begin{equation}\label{p}
V(\psi)=\frac{1}{2}\mu^2\psi^2,
\end{equation}
where $\mu>0$, and take the metric (\ref{6}) as
\begin{equation}\label{m1}
ds^2=-N(r)(1-\frac{2M(r)}{r})dt^2+(1-\frac{2M(r)}{r})^{-1}dr^2+r^2d\theta^2+r^2sin^2\theta d\phi^2.
\end{equation}
such that $M(r)$ is regular spherically symmetric mass function, satisfying $\lim_{r\to \infty}\frac{M(r)}{r}=0$.
For large $r$, the scalar field equation becomes
\begin{equation}\label{9}
\psi ^{\prime\prime}+\frac{2}{r}\psi ^{\prime}-\mu^{2}\psi =0.
\end{equation}
The solution of the above equation is
\begin{align}\label{10}
\psi(r)= B\frac{e^{\mu r}}{r}+C\frac{e^{-\mu r}}{r}.
\end{align}
In order to have a finite massive scalar field and also asymptotic flatness,
we choose $B=0$. Therefore
\begin{equation}\label{int}
\lim_{r\rightarrow\infty}\psi(r)=0.
\end{equation}

For the Dirichlet boundary condition, using Rolle's theorem, we find that there is an extremum point $r_0\in (r_s,\infty)$ such that $\psi'(r_0)=0$. So the Neumann and Dirichlet boundary conditions both require that the radial derivative of the scalar field be zero at a radius $r_0$ (for the Neumann boundary condition $r_0=r_s$).
By considering $\psi'(r_0)=0$, (\ref{7-1})and (\ref{7-2}) reduce to
\begin{eqnarray}\label{n3}
(rA'+A-1)\big|_{r_0}=-8\pi r^2 V(\psi(r_0))\nonumber \\
(rA'+A-1+\frac{N'}{N}Ar)\big|_{r_0}=-8\pi r^2 V(\psi(r_0)),
\end{eqnarray}
giving
\begin{equation}
N'(r_0)=0.
\end{equation}
By Using (\ref{m1}) we find
\begin{equation}
M'(r_0)= 4\pi r_0^2 V(\psi(r_0)).
\end{equation}

For a constant $M(r)$, $V(\psi(r_0))=0$ which for $V(\psi)=\frac{1}{2}\mu^2\psi^2$ gives $\psi(r_0)=0$.  Now as before by taking consecutive radial derivative of (\ref{7-4}) at $r_0$, and using (\ref{n3})and $\psi(r_0)=0$, we find that at each order  $\psi^{(j)}(r_0)=0$, so the only analytic well defined $\psi(r)$ is a trivial constant function, $\psi(r)=0$.

For a general $M(r)$ we proceed as follows:  Radial derivatives of (\ref{7-1}) and (\ref{7-2}) give
\begin{eqnarray}\label{f}
\left(rA''+2A'+16\pi r V=16\pi k r A^2\psi''^2\right)\big|_{r=r_0}\nonumber \\
\left(rA''+2A'+16\pi r V=-rA \frac{N''}{N}\right)\big|_{r=r_0}.
\end{eqnarray}
Therefore for $A(r)=1-\frac{2M(r)}{r}$ we have
\begin{equation}
\frac{N''}{N}\Big|_{r=r_0}=-16\pi k (1-\frac{2M(r_0)}{r_0})\psi''^2(r_0).
\end{equation}
For $N''(r_0)=0$ (which for example holds for a constant $N$) we have $\psi''^2(r_0)=0$, which from (\ref{7-4}) gives $\psi(r_0)=0$. So with the same argument used in the case of  constant $M$, and by using (\ref{7-4}), we find that $\psi$ and all of its derivatives vanish at $r_0$, hence it is a constant function $\psi(r)=0$. \newline
For $N''(r_0)\neq 0$, and For $k>(<)0$, there is no hairy solution for (\ref{m1}), with a convex (concave) $N$ at $r_0$. So for example, for $k>0$ we have not a hairy solution for (\ref{m1})with a convex $N$.

For $k<0$ the no scalar hair theorem holds for a Dirichlet reflecting star for a general $N$ and $M$. To see this we write (\ref{7-4}) at $r_0$ as
\begin{equation}\label{nb}
A(r_0)\left(kr_0A'(r_0)+k(A(r_0)-1)+r_0^2\right)\psi''(r_0)=r_0^2 V_{\psi}(r_0).
\end{equation}
Using (\ref{n3}) we obtain
\begin{equation}
\left(1-\frac{M(r_0)}{2r_0}\right)\left(-8\pi kr_0^2 V(\psi(r_0))+r_0^2\right)\psi(r)\psi''(r_0)=\mu^2 r_0^2\psi^2(r_0).
\end{equation}
But to have non trivial scalar field solution, at $r_0$ we must have $\psi(r_0)\psi''(r_0)<0$, which does not hold for a positive potential and $k<0$ .

\section{Conclusion}
In the non-minimal derivative coupling model (a model which was first proposed to describe the late time acceleration and also the Higgs inflation), a regular non-trivial scalar field may exist in the background of a black hole. We examined the same problem for a reflecting star having a boundary surface (instead of the absorbing event horizon of the black hole)on which the scalar field satisfies the Dirichlet or the Neumann boundary conditions.  We showed that in the potential-less case, the no-hair theorem holds for the Neumann boundary condition, and also for the Dirichlet boundary condition provided that the value of the scalar field at the boundary be the same as its asymptotic value. For the massive case, we chose the metric (\ref{m1}) and tried to derive the necessary conditions required to obtain a well defined regular scalar hair. We showed that when the nonminimal coupling constant is negative ($k<0$), we have no massive scalar hair with Dirichlet boundary condition (see (\ref{nb}). For $k>0$, by studying the consequence of boundary conditions on the scalar field solutions and the metric parameters $(M(r), N(r))$, we specified some necessary conditions required to get hair such that their violation leads to the no-hair theorem.

\end{document}